\documentclass{mn2e}  
\usepackage{amssymb} 
\input epsf.sty  
\newif\ifAMStwofonts  
\AMStwofontstrue  
  
\begin{document}  
  
\title{Extinction due to amorphous carbon grains in red quasars from the 
SDSS}  
\author[Czerny et al.]  
{B. Czerny$^1$,  J. Li$^2$, Z. Loska$^1$, R. Szczerba$^3$\\  
 $^1$Copernicus Astronomical Center, Bartycka 18, 00-716 Warsaw, Poland\\ 
 $^2$Great Neck North High School, Great Neck, NY, 11024, USA\\
 $^3$Copernicus Astronomical Center, Rabia\' nska 8, 87-100 Toru\' n, Poland\\} 
\maketitle  
\begin{abstract}  
We construct a quasar extinction curve based on the blue and red
composite quasar spectra of Richards et al. (2003) prepared from the
SDSS survey. This extinction curve does not show any traces of the
2200 \AA~ feature characteristic of the Interstellar Medium, and this
indicates that graphite grains are likely absent close to quasar
nuclei. The extinction is best modeled by AC amorphous carbon grains, 
assuming a
standard distribution of grain sizes ($p=3.5$) but slightly larger
minimum grain size ($a_{min} = 0.016 \mu$m) and lower maximum grain
size ($a_{min} = 0.12 \mu$m) than the respective canonical values for
the interstellar medium. The dust composition is thus similar to that
of the dust in AGB stars. Since graphite grains form from amorphous
carbon exposed to strong UV irradiation the results indicate that
either the dust forms surprisingly far from the active nucleus or in a
wind that leaves the nucleus quickly enough to avoid crystallization
into graphite.
  
\end{abstract}  
  
  
\section{Introduction}
\label{sect:intro} 

It is well known that the spectra of Seyfert galaxies -- particularly
Seyfert 2 galaxies - are considerably modified by circumnuclear
extinction.  It is also believed that Narrow Line Seyfert galaxies are
on average redder than typical Seyfert 1 galaxies because of reddening
due to the presence of dust (e.g.
Kuraszkiewicz \& Wilkes 2000, Constantin \& Shields 2003).  Evidence
is now accumulating that a significant fraction of bright AGN,
including high redshift quasars, are affected by dust.

The dust is seen directly in emission (IR continuum of dust and
emission lines of accompanying molecular gas; e.g. Priddey et
al. 2003, Omont et al. 1996, Kuraszkiewicz et al. 2003, Andreani,
Francheschini \& Granato 1999).  The presence of dust may also be
detected indirectly because of its effect on the quasar continuum in
the optical/UV band. Recent surveys show the widespread presence of
red quasars in quasar samples, and reddening by dust is the most
plausible explanation for their spectral shape (e.g. FIRST quasar
sample, White et al. 2003; SDSS quasar survey, Richards et al. 2003;
2MASS red quasars, Marble et al. 2003).

Knowledge of dust properties in the close vicinity of an active
nucleus is thus important both for determining intrinsic AGN spectra
and understanding the conditions in the gas surrounding the nucleus.

It has already been argued that the properties of the dust in such
extreme conditions need not be similar to the dust properties in the
interstellar medium of our Galaxy. It has been found that the
extinction curve of Pr\' evot et al.  (1984) determined for the Small
Magellanic Cloud (SMC) roughly applies to AGN dust (e.g. Czerny, Loska
\& Szczerba 1991, Laor \& Draine 1993, Pitman et al. 2000, Kuhn et
al. 2001, Richards et al.  2003).

In this paper we use the composite spectra of quasars from the Sloan
Digital Sky Survey (SDSS) prepared by Richards et al. (2003) to
determine an observational quasar extinction curve.  We then compare
this curve with model curves and argue that the circumnuclear regions
of AGN are filled with amorphous carbon grains. We provide a simple 
empirical formula for this new extinction law.
  
\section{Extinction curve from red SDSS quasars}
\label{sect:exti}

Results from the SDSS were used by Richards et al. (2003) to
create 6 composite quasar spectra (see their Fig. 7), selected
according to the adopted ranges of the (g-i) color excess. Composite
1, made from 770 objects, contains the bluest quasars in the
sample. Their optical/UV slope is well fitted by a power law, $F_{\nu}
\propto \nu^{\alpha}$, with energy index $\alpha = -0.25$. Composites
1 to 4 each consist of 770 objects.  Composite 5 consists of 211
objects well representing a red quasar
population; only objects with large color excesses were included, but
the few reddest objects were rejected in order to preserve the
completeness of the sample.

We assume that the bluest composite (composite 1) is essentially
unaffected by extinction (we will return to this point in the
Discussion). We then assume that other composites intrinsically
represent the same spectrum as composite 1, except affected by dust.

We can therefore directly construct the observed quasar extinction
curve for each of the composites from 2 to 5 vs. composite 1 by using
its usual definition,
\begin{equation}
 X(\lambda) = {A_{\lambda-V} \over E_{B-V}} = {A_{\lambda} - A_V \over A_B - A_V},
\label{eq:X}
\end{equation}
where 
\begin{equation}
A_{\lambda} = 2.5 \log (F_{\lambda}^1/F_{\lambda}^i)
\label{eq:alambda}.
\end{equation}
Here $F_{\lambda}^1$ is the composite 1 spectrum and $F_{\lambda}^i$
is one of the redder composites.  The values $A_V$ and $A_B$ are
calculated at the corresponding visual (5500 \AA) and blue (4400 \AA)
wavelengths.

\begin{figure}  
\epsfxsize = 90 mm  
\epsfbox{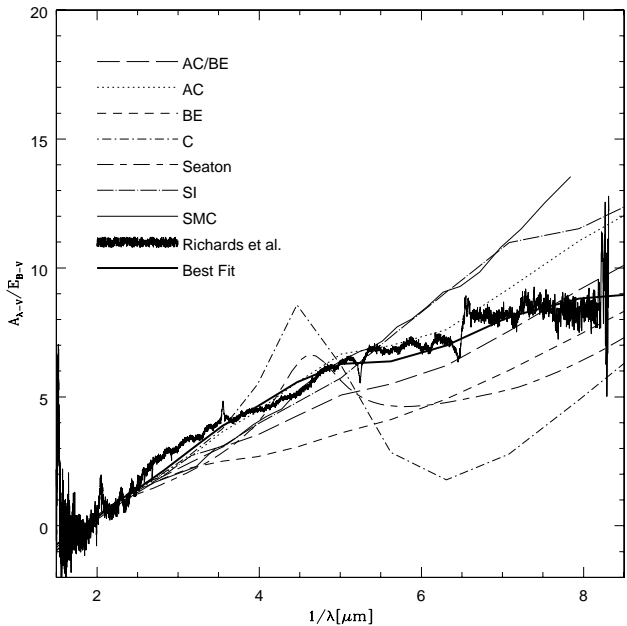}  
\caption{The quasar extinction curve (Richards et al. line) derived
assuming that composite 1 represents the intrinsic spectrum of all the
composites. Seaton and SMC lines show observationally determined extinction
curves in the Interstellar Medium (Seaton 1979) and in the Small Magellanic 
Cloud (Pr\' evot et al. 1984). 
The Best Fit line shows the favored model extinction
curve: AC amorphous carbon grains, $a_{min} = 0.016 \mu$m, $a_{max} =
0.12 \mu$m, $p = 3.5$. Other lines show typical model extinction
curves obtained assuming the grain distribution characteristic of the
interstellar medium ($a_{min} = 0.005 \mu$m, $a_{max} = 0.25 \mu$m, $p
= 3.5$) but for different materials (Si - silicates, AC - pure AC
amorphous carbon, BE - pure BE amorphous carbon, C - graphite).
\label{fig:ext1}}  
\end{figure}  
The curves obtained from composites 2 and 2n were very noisy since the
differences between the spectra were small, so we neglected them.  We
next averaged the curves obtained from composites 3, 4 and 5 in order
to reduce errors, thus creating a single curve.
 
The resulting extinction curve in the range covered by the SDSS data
is shown in Fig.~\ref{fig:ext1}. The curve is monotonic with
wavelength, without any trace of the 2200 \AA~ absorption feature. It
thus resembles the observationally determined extinction curve for
dust in the Small Magellanic Cloud (Pr\' evot et al. 1984) more than
the Seaton (1979) extinction curve that characterizes the interstellar
dust in our galaxy.  Accordingly, if the composite spectra are indeed
shaped by circumnuclear extinction, the dust composition close to
quasar nuclei must be non-standard.

In order to analyze the quasar extinction curve properties, we compare
it with several theoretical extinction curves.

\section{Theoretical extinction curves}
\label{sect:theo}

We construct the model extinction curves for the dust surrounding an
active nucleus following the approach of Czerny et al. (1995).

We assume that the nucleus is a source of radiation with a spectral
shape given by the sum of two components: a power law component
extending from IR to X-rays, and an accretion disc component
dominating the optical/UV part.  The power law component is
spherically symmetric while the disc component intensity depends on
the cosine of the inclination angle to the disc symmetry axis.
  
The innermost part of an AGN is surrounded by a spherical dust
shell. The inner radius of the dusty shell is determined by the
temperature of dust evaporation. The outer radius of the shell is
assumed to be at 300 pc, and the adopted density profile is given in
the form of a power law, $n(r) \propto r^{-0.75}$, the favored one by
models of the far-IR spectra of AGN (e.g. Loska, Szczerba \& Czerny
1993, Granato, Danese \& Franceschini 1997). The results are not very
sensitive to these parameters so long as the spatial distribution is
rather flat.

We determine the theoretical extinction curve by solving for the
radiation transfer through the dust surrounding the
envelope, determining the dust-reddened radiation spectrum, and
comparing this spectrum to the intrinsic spectrum of the source, using
Eqs.~\ref{eq:X} and \ref{eq:alambda}.  The radiative transfer is
solved for as described by Loska et al. (1993). The temperature of the
grains of a given size and chemical composition is calculated from the
energy balance. We assume a maximum carbon grain temperature of 1500 K
and maximum silicate dust temperature of 1000 K, and we allow for
selective grain evaporation close to the inner edge of the dusty
cloud. The radiative transfer is solved for assuming that the dust is
optically thin, i.e. we neglect multiple scattering effects.

Optical properties of astronomical silicate and carbon grains are
taken from Draine (see Laor \& Draine 1993), and the amorphous carbon
properties were calculated using Mie theory and the optical constants of
Martin (see Rouleau \& Martin 1991).  Some modifications were
introduced for large grains/short wavelengths (see Czerny et al. 1995
for details). The distribution of grain sizes was assumed to be in
a power law form, $n(a) \propto a^{-p}$, with minimum and maximum
grain sizes given by $a_{min}$ and $a_{max}$.  The standard values of
these parameters in the ISM are: $p = 3.5$, $a_{min} = 0.005 \mu$m and
$a_{max} = 0.25 \mu$m (Mathis, Rumpl \& Nordsieck 1977).  We generally
tried to minimize departure from these values.

The extinction curves representing various chemical compositions of
dust grains but otherwise based on the dust parameters $p, a_{min}$,
and $a_{max}$ of the ISM are shown in Fig~\ref{fig:ext1}. The observed
extinction curve roughly follows at long wavelengths the extinction
curve of dust consisting of AC amorphous carbon grains. The departure
is seen only at short wavelengths above $\sim 1500 $ \AA.

We can find a better representation of the observed curve assuming AC
composition but allowing for a change in the minimum and maximum grain
sizes.  In Fig.~\ref{fig:ext1} we show such a curve with $a_{min} =
0.016 \mu$m and $a_{max} = 0.12 \mu$m.

We also analyzed the properties of dust of other chemical
compositions, but they did not mimic the quasar extinction curve
equally well. In particular, no other standard or non-standard
extinction curve could reasonably trace the observational extinction
curve in the $.2\mu$m to $.5 \mu$m range.  Furthermore, we do not
expect large amounts of silicates in AGN surroundings; silicate
emission features expected in the IR are weak or absent in most AGN
(Roche et al. 1991), and efficient formation of CO molecules offers a
natural explanation for the depletion of the oxygen necessary for
silicate formation (Rawlings \& Wiliams 1989).  Any significant
contribution from graphite also resulted in the presence of a 2200
\AA~ peak and no traces of such a peak are in the data.  We thus argue
that AC amorphous carbon grains best describe the observational
extinction curve based on the Richards et al. (2003) data because there are
significant errors when attempting to model the curve based on any
other dust grains.

\section{Discussion}
\label{sect:diss}

The lack of the 2200 \AA~ absorption feature due to graphite grains 
characteristic of interstellar extinction was already noted in
several papers concerning AGN (e.g. Pitman et al. 2000, 
Kuhn et al. 2001, Richards et al. 2003) and
therefore an SMC extinction curve seems to apply better than the Seaton 
ISM extinction
curve. Our results based on SDSS quasars from Richards et al. (2003)
support this view.   

Our best model of the observational quasar extinction curve is made
from amorphous AC carbon grains. We assumed a standard power law size
distribution of grains, $p = 3.5$, but better agreement with the
observational curve required a narrower size range (minimum size $0.016
\mu$m, maximum size $0.12 \mu$m) than the canonical one of Mathis
et al. (1977), 0.005 and 0.25$\mu$m.

This extinction curve is slightly flatter than the SMC extinction
curve at short wavelengths. The reddening law corresponding to it can
be well approximated by the following empirical formula
\begin{equation}
{A_{\lambda} \over E_{B-V}} = -1.36 + 13 \log(1/\lambda) [\mu m],
\label{eq:new}
\end{equation}
in the range of $1/\lambda$ between 1.5 and 8.5 $\mu$m$^{-1}$.

Our analysis was based on an average of several observational curves
derived from Richards et al. (2003) quasar data.  We derived
extinction curves assuming that their composite 1 was the intrinsic
spectrum of all the composites.  We were then able to construct
extinction curves based on composite 1 vs. composites 2 through 5 and
composite 1n vs. composites 2n through 5n.  We ended up disregarding
the composite 1 vs. composite 2 and composite 1n vs. composite 2n
extinction curves because they were too noisy.  We then averaged the
extinction curves based on composite 1 vs. composites 3 through 5.  We
selected the composite 1 vs. composite 5 extinction curve over the
composite 1n vs. composite 5n curve because the former has higher
signal-to-noise.  The latter was corrected for redshift and magnitude
differences, but we found that the two curves were quite similar with
the latter having more noise.  We did not include the 2n through 5n
subgroups in our average because these quasars were already included
in the 2 through 5 composites.  We found that the composite 1
vs. composites 3 through 5 and composite 1n vs. composites 3n through
5n created a small range of possible extinction curves.  We believe
that the true observational extinction curve lies somewhere within
this range, and so our final observational extinction curve is an
average of the curves that are contained in this range.  We find that
AC amorphous carbon grains are able to model the extinction curves
within this range as well as they model the average extinction curve.
We also find that even the lower extinction curves in this range are
modeled with difficulty when taking into account other types of dust
grains.

Our analysis used the composite spectra. Any such spectra, including
those of Richards et al. (2003), contain systematical errors which are
difficult to assign. The errors come partially from the calibration 
uncertainty, and partially from the systematic differences in the 
objects constituting the short wavelength and the long wavelength parts
of the final spectrum. Therefore, determination of the errors of the
derived extinction curve is rather difficult. Although the conclusion 
about the dominating role of amorphous carbon
seems to be firm, the grain size range needs confirmation.

This dust composition is not surprising taking into account that the
conditions around an active nucleus are quite similar to conditions in
the envelopes of asymptotic giant branch (AGB) stars. Radiation
flux, pressure and densities may be comparable (see e.g. Elvis,
Marengo and Karovska 2002), particularly taking into account that the
cool gas in AGN is possibly in the form of clouds.  This is indicated by
the presence of the Broad Line Region and Narrow Line Region implied
by AGN spectra.  Extinction in carbon-rich AGB stars is also dominated
by amorphous carbon, although in some stars with additional silicon
carbide (e.g. Piovan, Tantalo \& Chiosi 2003).

The 2200 \AA~ feature seen in the interstellar medium is caused by
single graphite and/or clusters of graphite grains (see e.g. Draine \&
Lee 1984, Andersen et al. 2003).  Laboratory data indicate that
graphite grains form from amorphous carbon grains as a result of
exposure to strong UV radiation (Ogmen \& Duley 1988; see also
Dorschner \& Henning 1995). AGN environments provide intensive UV
flux, much more than the relatively cool environments of AGB stars
(their temperature is of the order of 2000 - 4000 K), yet graphite
grains apparently do not form in AGN surroundings.  
This may support the view of
Elvis et al. (2002) that the dust forms in an outflowing wind. The
dust forms and is exposed to strong UV radiation only for a time period
of the order of years to hundreds of years, most probably not enough
time to form the regular crystal structures of graphite.
Alternatively, dust may form relatively far from the nucleus. However,
time delay measurements between the optical and IR band indicate hot
dust distances not larger than $\sim 1$ pc for bright objects
(e.g. Sitko et al. 1993, Ulrich, Maraschi \& Urry 1997).

The composite 1 spectrum is most probably only weakly reddened, if at
all.  We estimate this in the following way. We assume that the
amorphous carbon grain extinction curve also well represents the
reddening within the composite 1 objects. We allow for an arbitrary
amount of reddening and compare the resulting spectra with the
standard accretion disc spectra of Shakura \& Sunyaev (1973). Such
models represent bright and blue quasar spectra quite well (see
e.g. Koratkar \& Blaes 1999 for the discussion of the quasar composite
spectrum of Francis et al. 1991).

\begin{figure}  
\epsfxsize = 90 mm  
\epsfbox{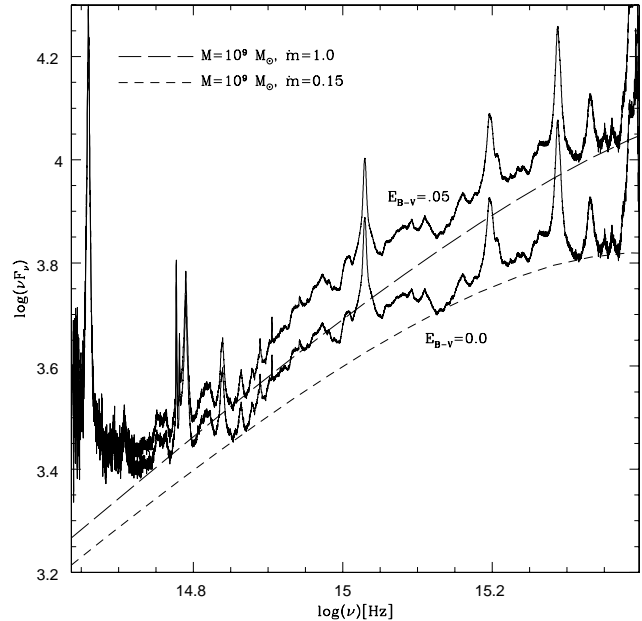}  
\caption{The composite 1 of Richards et al. (2003) (lower spectrum)
and the same spectrum, but dereddened by $E_{B-V}=0.05$ using our AC
amorphous extinction curve (upper spectrum).  Underlying
continua of both spectra can be represented by the the disc spectra
models assuming different accretion rates. 
\label{fig:disk}}  
\end{figure}

We can model the composite 1 spectrum assuming any value of $E_{B-V}$
between 0 and $\sim 0.05$ by changing the mass and accretion rate of
our model. Reddening larger than $E_{B-V} = 0.05 $ is excluded since
it leads to an intrinsic spectrum harder than the asymptotic
energy index value of 0.33 predicted by the disc model. 
These two extreme cases of $E_{B-V} = 0$ and 0.05 
are shown in Fig.~\ref{fig:disk}.  For all intermediate reddening
values the model well represents the overall continuum.  However,
there is evidently some flattening of the observed spectrum towards IR
(see Fig.~\ref{fig:disk}). This effect is well known in AGN although
its nature is not clear. It might be caused by the contribution of
some starlight in the host galaxy (clearly important for less luminous
AGN although not necessarily for bright quasars, see e.g. Elvis et
al. 1994), the illumination of the outer disc by the radiation
generated in its innermost region, or perhaps some contribution of
non-thermal emission to the overall spectrum.

The good correspondence between the dereddened spectrum of the bright
quasars with a simple accretion disc model indicates that in these
sources the production of X-ray emission does not modify an optically
thick flow. In contrast to much fainter Seyfert 1 galaxies, X-ray
emission in quasars must therefore make up only a small fraction of
the bolometric luminosity. Studies of broad band $\alpha_{ox}$ index
trends with luminosity indeed seem to support such a view
(e.g. Bechtold et al. 2003).

\section{Conclusions}

Analysis of SDSS quasar composites shows that quasar spectra are
affected by circumnuclear dust composed primarily of 
AC amorphous carbon grains. The derived observational extinction curve is 
similar to SMC extinction but flatter at the shortest wavelengths.
Eq.~\ref{eq:new} provides a simple empirical formula for the derived quasar
extinction law.
 
\section*{Acknowledgements}  
 
This work was supported in part by grants 2P03D~003~22 (BCz \& ZL)
of the Polish State Committee for Scientific Research (KBN). JL
acknowledges financial support from the Institute of Physics,
Polish Academy of Sciences.

\ \\  
This paper has been processed by the authors using the Blackwell  
Scientific Publications \LaTeX\  style file.  
 
\end{document}